\documentclass[a4paper,twocolumn,showpacs5]{revtex4}
\usepackage[english]{babel}
\usepackage{graphicx}
\usepackage{xspace}
\usepackage{longtable}
\usepackage{amsmath,amssymb}

\DeclareMathAlphabet{\mathbfit}{OT1}{cmr}{bx}{it}

\begin{document}
\title{Structural and doping effects in the half-metallic double perovskite $A_2$CrWO$_6$}

\author{J.~B.~Philipp}
\email{Boris.Philipp@wmi.badw.de}
\affiliation{Walther-Mei{\ss}ner-Institut, Bayerische Akademie der
Wissenschaften, Walther-Mei{\ss}ner Str.~8, 85748 Garching, Germany}
\author{P.~Majewski}
\affiliation{Walther-Mei{\ss}ner-Institut, Bayerische Akademie der
Wissenschaften, Walther-Mei{\ss}ner Str.~8, 85748 Garching,
Germany}
\author{L.~Alff}
\email{Lambert.Alff@wmi.badw.de}
\affiliation{Walther-Mei{\ss}ner-Institut, Bayerische Akademie der
Wissenschaften, Walther-Mei{\ss}ner Str.~8, 85748 Garching, Germany}
\author{A.~Erb}
\affiliation{Walther-Mei{\ss}ner-Institut, Bayerische Akademie der
Wissenschaften, Walther-Mei{\ss}ner Str.~8, 85748 Garching, Germany}
\author{R.~Gross}
\affiliation{Walther-Mei{\ss}ner-Institut, Bayerische Akademie der
Wissenschaften, Walther-Mei{\ss}ner Str.~8, 85748 Garching, Germany}
\author{T.~Graf}
\affiliation{Walter-Schottky-Institut, TU M\"{u}nchen, Am Coulombwall
3, 85748 Garching, Germany}
\author{M.~S.~Brandt}
\affiliation{Walter-Schottky-Institut, TU M\"{u}nchen, Am Coulombwall
3, 85748 Garching, Germany}
\author{J. Simon}
\affiliation{Institut f\"{u}r Anorganische Chemie, Universit\"{a}t Bonn,
R\"{o}merstr. 164, 53117 Bonn, Germany}
\author{T. Walther}
\affiliation{Institut f\"{u}r Anorganische Chemie, Universit\"{a}t Bonn,
R\"{o}merstr. 164, 53117 Bonn, Germany}
\author{W. Mader}
\affiliation{Institut f\"{u}r Anorganische Chemie, Universit\"{a}t Bonn,
R\"{o}merstr. 164, 53117 Bonn, Germany}
\author{D.~Topwal}
\affiliation{Solid State and Structural Chemistry Unit, Indian
Institute of Science, Bangalore 560 012, India}
\author{D.~D.~Sarma}
\affiliation{Solid State and Structural Chemistry Unit, Indian
Institute of Science, Bangalore 560 012, India}

\date{\today}

\begin{abstract}
The structural, transport, magnetic and optical properties of the double
perovskite $A_2$CrWO$_6$ with $A=\text{Sr, Ba, Ca}$ have been studied. By
varying the alkaline earth ion on the $A$ site, the influence of steric
effects on the Curie temperature $T_C$ and the saturation magnetization has
been determined. A maximum $T_C=458$\,K was found for Sr$_2$CrWO$_6$ having
an almost undistorted perovskite structure with a tolerance factor $f\simeq
1$. For Ca$_2$CrWO$_6$ and Ba$_2$CrWO$_6$ structural changes result in a
strong reduction of $T_C$. Our study strongly suggests that for the double
perovskites in general an optimum $T_C$ is achieved only for $f \simeq 1$,
that is, for an undistorted perovskite structure. Electron doping in
Sr$_2$CrWO$_6$ by a partial substitution of Sr$^{2+}$ by La$^{3+}$ was
found to reduce both $T_C$ and the saturation magnetization $M_s$. The
reduction of $M_s$ could be attributed both to band structure effects and
the Cr/W antisites induced by doping. Band structure calculations for
Sr$_2$CrWO$_6$ predict an energy gap in the spin-up band, but a finite
density of states for the spin-down band. The predictions of the band
structure calculation are consistent with our optical measurements. Our
experimental results support the presence of a kinetic energy driven
mechanism in $A_2$CrWO$_6$, where ferromagnetism is stabilized by a
hybridization of states of the nonmagnetic W-site positioned in between the
high spin Cr-sites.
\end{abstract}
\maketitle

\section{Introduction}

The investigation of ordered double perovskite materials $A_2 BB^\prime
\text{O}_6$ with $A$ an alkaline earth such as Sr, Ba, or Ca and
$B,B^\prime$ two different transition metals has been strongly stimulated
by the discovery of a large room temperature magnetoresistive effect at low
magnetic fields in Sr$_2$FeMoO$_6$ \cite{Kobayashi:98}. The fact that the
double perovskites seem to be ferromagnetic metals with high Curie
temperatures $T_C$ of up to 635\,K \cite{Kato:02} and apparently have
highly spin polarized conduction band makes these materials interesting for
applications in spintronic devices such as magnetic tunnel junctions or
low-field magnetoresistive sensors \cite{Wolf:2001}. However, the double
perovskites are also of fundamental interest since both their basic physics
and materials aspects are not well understood.

It is evident that intensive research has been dedicated to both the
variation of the metallic/magnetic ions on the $B$- and $B^\prime$-site as
well as electron doping studies, where the divalent alkaline earth ions on
the $A$-site are partially replaced by a trivalent rare earth ion such as
La, in order to understand the electronic structure and the magnetic
exchange in the double perovskites in detail. Furthermore, these studies
aimed for the tailoring and optimization of the magnetic properties of the
double perovskites for their use in magnetoelectronic devices such as spin
valves, magnetic information storage systems, or as sources for spin
polarized electrons in spintronics. Here, important aspects are the
achievement of sufficiently high values for $T_C$ and the spin polarization
to allow for the operation of potential devices at room temperature. Along
this line, in the compound Sr$_2$CrReO$_6$ a Curie temperature of $T_C
=635$\,K has been obtained \cite{Kato:02}. Furthermore, in Sr$_2$FeMoO$_6$
an increase in $T_C$ of about 70\,K has been reported as a result of
electron doping by partial substitution of divalent Sr$^{2+}$ by trivalent
La$^{3+}$ \cite{Navarro:01}. With respect to magnetoresistance, up to now
large low-field magnetoresistive effects have been found not only in
Sr$_2$FeMoO$_6$\cite{Kobayashi:98,Sarma:00a,Garcia:01}, but also in many
other double perovskites as for example Sr$_2$FeReO$_6$
\cite{Kobayashi:99}, Sr$_2$CrWO$_6$ \cite{Philipp:01}, and
(Ba$_{0.8}$Sr$_{0.2}$)$_{2-x}$La$_x$FeMoO$_6$ \cite{Serrate:02}.

The origin of magnetism in the double perovskite still is discussed
controversially. Historically, the ferrimagnetism in the system
Sr$_2$FeMoO$_6$ has been explained in terms of an antiferromagnetic
superexchange interaction between the Mo$^{5+}$ ($5d^1$) spin and the
Fe$^{3+}$ ($3d^5$) spins \cite{Galasso:1966,Patterson:1963,Longo:1961}.
However, more recently Moritomo {\it et al.} found a strong correlation
between the room temperature conductivity and the Curie temperature,
implying that the mobile conduction electrons mediate the exchange
interaction between the local Fe$^{3+}$ spins
\cite{Moritomo:00,Moritomo:00e}. Therefore, in analogy to the doped
manganites it was tempting to explain the ferromagnetic coupling between
the Fe sites based on a double exchange mechanism, where the delocalized
electron provided by the Mo~$4d^1$ configuration plays the role of the
delocalized $e_g$ electron in the manganites. However, as pointed out by
Sarma~\cite{Sarma:00} there are important differences between the
manganites and the double perovskites. In the former, both the localized
Mn~$t_{2g}$ electrons and the delocalized Mn~$e_g$ electron reside at the
same site and their spins are coupled ferromagnetically by a strong on-site
Hund's coupling. In the latter, the localized Fe~$3d$ electrons (Fe$^{3+}$:
$3d^5, S=5/2$) and the delocalized Mo~$4d$ electron (Mo$^{5+}$: $4d^1,
S=1/2$) nominally are at two different sites although the Mo~$4d$ electron
obtains a finite Fe character by sizable hopping interaction. At first
glance this seems to support a double exchange szenario. However, in
Sr$_2$FeMoO$_6$ according to band structure calculations the Fe~$3d$
spin-up band is completely full making it impossible for another spin-up
electron to hop between Fe sites and thus forcing the delocalized electrons
to be spin-down electrons. Therefore, the Hund's coupling strength, which
provides the energy scale for the on-site spin coupling in the double
exchange mechanism for the manganites, cannot be invoked for
Sr$_2$FeMoO$_6$. This demonstrates that in the double perovskites the
antiferromagnetic coupling between the localized and the delocalized
electrons, which according to the large $T_C$ is strong, must originate
from another mechanism. Such mechanism has been proposed by
Sarma~\cite{Sarma:00} for Sr$_2$FeMoO$_6$ and extended to many other
systems by Kanamori and Terakura~\cite{Kanamori:01,Fang:01}. In this model,
the hybridization of the Mo~$4d$ ($t_{2g}$) and Fe $3d$ ($t_{2g}$) states
plays the key role in stabilizing ferromagnetism at high Curie temperatures
\cite{Sarma:00,Kanamori:01,Fang:01}. If the Fe$^{3+}$ spins are
ferromagnetically ordered, the hybridization between the Fe~$3d$ ($t_{2g}$)
and the Mo~$4d$ ($t_{2g}$) states pushes up and down the Mo~$4d_\uparrow$
and Mo~$4d_\downarrow$ states, respectively. The essential point in this
scenario, which is discussed in more detail below, is that the hybridized
states are located energetically between the exchange-split
Fe~$3d_\uparrow$ and the Fe~$3d_\downarrow$ levels. We note that in this
model the magnetic moment at the Mo site is merely induced by the Fe
magnetic moments through the hybridization between the Fe~$3d$ and the
Mo~$4d$ states what can be considered as a magnetic proximity effect. In
this sense the double perovskites should be denoted ferromagnetic and not
ferrimagnetic~\cite{Tovar:2002}.

Despite the recent progress in understanding the physics of the double
perovskites there is still an open debate on the adequate theoretical
modelling. In particular, the details of the interplay between structural,
electronic, and magnetic degrees of freedom in the double perovskites is
not yet clearly understood. Recently, the $T_C$ in double perovskites was
discussed to depend sensitively on the band structure and band filling in
contrast to experimental results \cite{Phillips:03}. However, since in
experiments often changes in the number of conduction electrons achieved by
partially replacing the divalent alkaline earth ions on the $A$- and
$A^\prime$-site by trivalent rare earth ions are accompanied by significant
steric effects, it is difficult to distinguish between doping and
structural effects. It is therefore desirable to investigate both the
influence of structural changes (changes in the bond length and bond
angles) and the effect of carrier doping.

Here, we present a study of the effect of structural changes and doping in
the double perovskite $A_2$CrWO$_6$ ($A = \text{Sr, Ba, Ca}$) on the
magnetotransport, the magnetic and optical properties together with band
structure calculations. We note that a major difference between the system
$A_2$CrWO$_6$ and the systems $A_2$FeMoO$_6$ or $A_2$FeReO$_6$ is the fact
that in the former the majority spin band is only partially filled
(Cr$^{3+}$, $3d^3, S=3/2$, only the Cr~$3d$ ($t_{2g}$) levels are
occupied), whereas in the latter the majority spin band is completely
filled (Fe$^{3+}$, $3d^5, S=5/2$, both the Fe~$3d$ ($t_{2g}$) and Fe~$3d$
($e_{g}$) levels are occupied). We discuss that despite this difference the
magnetism in $A_2$CrWO$_6$ and $A_2$FeMoO$_6$ is similar. For both systems
the exchange gap is important for the magnetic exchange and half
metallicity. However, whereas for the former system also the crystal field
gap between the $t_{2g}$ and the $e_g$ levels plays a key role, for the
latter the crystal field gap is irrelevant. A key result of our study is
that optimum magnetic properties such as half metallicity and high $T_C$
can only be obtained in the undoped compounds close to the ideal
undistorted perovskite structure that is characterized by a tolerance
factor $f\simeq 1$ of the perovskite unit cell. Furthermore, we find that
electron doping tends to decrease $T_C$ in $A_2$CrWO$_6$.

\section{Sample Preparation}

In our study, we have used both polycrystalline and epitaxial thin film
samples. Polycrystalline samples were prepared from stoichiometric mixtures
of SrCO$_3$, BaCO$_3$, CaCO$_3$, Cr$_2$O$_3$, La$_2$O$_3$, and WO$_3$ with
a purity ranging between 99.99\% and 99.999\%. These powders were
thoroughly mixed, placed in Al$_2$O$_3$ crucibles, and then repeatedly
heated in a thermobalance under reducing atmosphere (H$_2$/Ar: 5\%/95\%)
with intermediate grinding. The final sintering temperatures $T_{\rm sint}$
were increased from 1200$^\circ$C for the first up to 1550$^\circ$C for the
final firing for the first series of samples. In order to increase the
Curie temperature we prepared a second series of undoped Sr$_2$CrWO$_6$
samples with more sintering steps and longer final sintering times at lower
(1300$^\circ$C) temperatures. For this series we find the highest Curie
temperatures, which correspond to the values reported in literature. This
most likely is caused by a slightly higher oxygen content in these samples
compared to those sintered at higher temperatures in the same reducing
atmosphere.  The use of the thermobalance allowed to monitor the ongoing
reaction process due to the associated weight loss of the samples. The
exact oxygen content of the investigated samples has not been determined.
However, due to the very similar preparation procedure for the samples used
for the study of electron doping, we can assume a similar oxygen content.

The polycrystalline samples were characterized by x-ray powder
diffractometry to detect parasitic phases, e.g. the insulating compound
SrWO$_4$. Interestingly, even in samples containing SrWO$_4$ no chromium
containing parasitic phases could be detected. The most simple explanation
for this observation is the high vapor pressure of Cr$_2$O$_3$ resulting in
a loss of chromium. However, no weight loss has been found in the
thermogravimetric measurement. Therefore, it is more likely that the
formation of the mixed phase $A_2$Cr$_{1+x}$W$_{1-x}$O$_{6\pm \delta}$
compensates for the missing Sr and W due to the SrWO$_4$ impurity phase.
While the double perovskite Sr$_2$FeMoO$_6$ can be grown as single crystal
\cite{Tomioka:00,Moritomo:00b,Moritomo:00c}, our attempts to grow single
crystals of Sr$_2$CrWO$_6$ by floating zone melting failed so far due to
the high vapor pressure of Cr$_2$O$_3$ at the melting point.

\begin{figure}[tb]
\centering{%
\includegraphics [width=0.8\columnwidth,clip=]{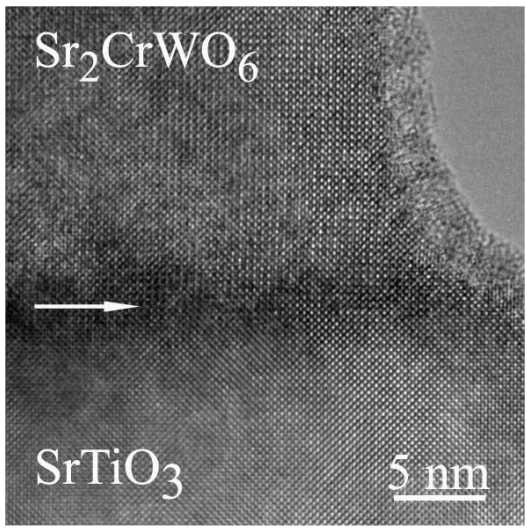}}
 \caption{Transmission electron micrograph of an epitaxial
Sr$_2$CrWO$_6$ thin film deposited on a SrTiO$_3$ substrate with low
lattice mismatch ($\simeq 0.13$\%). The image was taken along the $[010]$
direction.}
 \label{TEM}
\end{figure}

Epitaxial thin films of Sr$_2$CrWO$_6$ were fabricated in a UHV Laser
molecular beam epitaxy (L-MBE) system \cite{Gross:2000a}. The details of
the thin film deposition have been described recently
\cite{Philipp:01,Philipp:03}. Fig.~\ref{TEM} shows a transmission electron
micrograph of an epitaxial Sr$_2$CrWO$_6$ thin film grown on a SrTiO$_3$
(001) substrate. The image shows that the thin film grows epitaxially with
high crystalline quality. The $c$-axis oriented films have been analyzed
using a high resolution 4-circle x-ray diffractometer. Only ($00\ell$)
peaks could be detected and typically, rocking curves of the (004) peak had
a full width at half maximum (FWHM) of only 0.02 to 0.03$^{\circ}$, which
is very close to the FWHM of the substrate peak. Moreover, an AFM analysis
of the Sr$_2$CrWO$_6$ films showed that they have a very small surface
roughness of the order of one unit cell \cite{Philipp:03}. In this study
the thin films were used for optical measurements in order to determine the
band structure.

\section{Experimental Results and Discussion}
\subsection{Structural Properties}

It is well known that the tolerance factor $f$ determines the crystal
structure of perovskites $AB$O$_3$ \cite{Goodenough:rev}. Only for $f$
close to unity a cubic perovskite structure is obtained. For $f\ne 1$ a
tilt and rotation of the oxygen octahedra is obtained compensating for the
misfit of the ionic radii of the involved $A$ and $B$ cations. Hence, the
deviation of the tolerance factor from the ideal value $f=1$ can be used as
a measure for the internal strain in perovskites induced by the different
radii of the $A$ and $B$ cations. This can be seen from the definition of
$f$ given by
$$ f=\frac{ r_A + r_O }{\sqrt{2} \left(\langle r_B \rangle + r_O \right)}.$$
Here, $\langle r_B \rangle$ denotes the average ionic radius for the ions
on the $B$-site. For $f<1$, the strain is compensated by a tilt and
rotation of the oxygen octahedra. This results in a deviation of the
$B$-O-$B$ bond angles from the ideal value of $180^\circ$. For $0.96\leq
f\leq1$ the connecting pattern of the oxygen octahedra is rhombohedral,
whereas it is orthorhombic for lower values of $f$ \cite{Tokura:99}. A
simple consequence of the deviation of the bond angles from $180^\circ$ is
a decrease in the hopping amplitude because the electron transfer between
the $B$ sites is via O~$2p$-states. This, in turn, results in a decrease of
the one-electron bandwidth $W$. For $f>1.06$, a hexagonal structure is
expected which is classified by the stacking sequence of the $B$O$_6$
octahedra \cite{Shikano:95}.

\begin{table}[tb]
 \caption{Overview on the crystallographic properties of the investigated
samples of the $A_2$CrWO$_6$ series with $A=\text{Sr, Ba, Ca}$.
The tolerance $f$ factor was calculated using the SPuDS simulation
software \cite{Lufaso:01}.}
 \label{Table:Samples}
\begin{ruledtabular}
\begin{tabular}{c c c c}
  material       & structure                      & $f$ & parasitic \\
  $T_{\rm sint}$ & bond angle, lattice parameters &     &   phases     \\ \hline\hline
  Ca$_2$CrWO$_6$ & $P2_1/n$, $\beta=90.1^\circ$, $a=5.39$\,{\AA} & $0.945$ & CaWO$_4$ \\
   $1400^\circ$C & $b= 5.45$\,{\AA}, $c=7.66$\,{\AA} & & \\
                 & $a_{\text{pseudocubic}}=7.6 6$\,{\AA}        &         &               \\
                 & antisites Cr/W: 13\% &  & \\ \hline
  Sr$_2$CrWO$_6$ & $Fm\overline{3}m$, $a=7.82$\,{\AA} & $0.999$ & \\
  $1550^\circ$C  & antisites Cr/W: 23\% &  & SrWO$_4$ \\
  $1300^\circ$C  & antisites Cr/W: 31\% &  & Sr$_2$WO$_5$ \\
  \hline
  Ba$_2$CrWO$_6$ & $P\overline{6}2c$, $a=5.70$\,{\AA}, $c=13.99$\,{\AA} & $1.059$ & Ba$_3$WO$_6$ \\
   $1450^\circ$C & $a_{\text{pseudocubic}}=8.06$\,{\AA}         &         &               \\
                 & $B_1$ sites: 75\% Cr and 25\% W&         &               \\
                 & $B_2$ sites: 100\% W &         &               \\
\end{tabular}
\end{ruledtabular}
\end{table}

In the following, we discuss the change of tolerance factor $f$ and its
influence on the structure for the system $A_2$CrWO$_6$ with $A=\text{Sr,
Ba, Ca}$. An overview on the crystallographic properties of the
investigated samples is given in Table~\ref{Table:Samples}. We note that
samples with mixed contents of different alkaline earth ions on the
$A$-site had a strong tendency to phase separation and are therefore
excluded from our analysis. The value of $f$ was calculated using the bond
valence parameters as obtained using the SPuDS simulation software
\cite{Lufaso:01}. While Sr$_2$CrWO$_6$ has an almost ideal tolerance factor
$f\simeq 1$, for Ca$_2$CrWO$_6$ the tolerance factor is much smaller
($f=0.945$), and much larger than 1 ($f=1.059$) for Ba$_2$CrWO$_6$. The
crystal structure of the different compounds was determined by a Rietveld
refinement of the x-ray data. The result of the refinement is listed in
Table~\ref{Table:Samples}. As expected, the Sr$_2$CrWO$_6$ compound with
$f=0.999$ is cubic. In contrast, the Ca$_2$CrWO$_6$ compound with $f =
0.945$ is strongly distorted forming a monoclinic system as predicted by
the SPuDS program \cite{Lufaso:01}. The Ba$_2$CrWO$_6$ compound with
$f=1.059$ has a 6-layered hexagonal structure. The pseudocubic lattice
parameter for Ba$_2$CrWO$_6$ is about $8.06$\,{\AA}. The crystal structure is
identical to that of the compound Ba$_3$Cr$_2$WO$_9$, where the $B$ sites
are not equivalent. Two third of the sites are centered within face-shared
octahedra ($B_1$ site), whereas one third is centered within corner-shared
octahedra ($B_2$ site) \cite{Shikano:95}.

Antisite defects can occur if Cr and W ions exchange their positions on the
$B$ and $B^\prime$ sublattices. If Cr and W are randomly distributed,
antisites are 50\%. The amount of Cr/W antisites has been determined by a
Rietveld refinement of the x-ray data.  We first discuss the Sr$_2$CrWO$_6$
compound for which the amount of antisites is 23\%. This value is higher
than that obtained for the Fe/Mo system \cite{Garcia:01}. This most likely
is caused by the much smaller difference in the ionic radii
\cite{Shannon:76} between the Cr ($r_{\text{Cr$^{3+}$}} =0.615$\,{\AA}) and W
($r_{\text{W$^{5+}$}} = 0.62$\,{\AA}) ions compared to the significantly larger
difference between the Fe ($r_{\text{Fe$^{3+}$, high spin}} = 0.645$\,{\AA})
and Mo ($r_{\text{Mo$^{5+}$}} = 0.61$\,\AA) ions. In general, the amount of
antisites on the $B$ and $B^\prime$ sublattices is reduced by both large
differences in the ionic radius and the valence state. Therefore, in the
Cr$^{2+}$/W$^{6+}$ system ($r_{\text{Cr$^{2+}$}} = 0.73$\,{\AA}\ and
$r_{\text{W$^{6+}$}} = 0.60$\,{\AA}) the amount of antisites is reduced
compared to the Cr$^{3+}$/W$^{5+}$ system. The measured small amount (23\%)
of antisites in Sr$_2$CrWO$_6$ suggests an intermediate valence state.

The amount of Cr/W antisites in the Ca$_2$CrWO$_6$ compound was determined
to about 13\%. The lower amount of antisites in Ca$_2$CrWO$_6$ (see
Table~\ref{Table:Samples}) may be associated with a gradual transition from
W$^{5+}$ to W$^{6+}$. The situation for the Ba$_2$CrWO$_6$ compound is more
difficult.  While in Ba$_3$Cr$_2$WO$_9$ the Cr and W ions each are located
only on one sort of $B$ site, this is not the case for Ba$_2$CrWO$_6$. The
Rietveld refinement shows that the W ions are occupying almost completely
the $B_2$ sites. The rest of the W ions and the Cr ions is located on the
$B_1$ sites. A possible order of W and Cr ions on the $B_1$ sites was not
considered in our analysis.

We briefly compare the crystallographic properties of the $A_2$CrWO$_6$
systems to those of other double perovskites.  For the system
$A_2$FeMoO$_6$ the use of different alkaline earth ions on the $A$ site
results in the following crystal structures: Ca$_2$FeMoO$_6$ with $f=0.954$
is monoclinic or orthorhombic, Sr$_2$FeMoO$_6$ with $f=1.009$ is about
cubic (with very small tetragonal distortion), and Ba$_2$FeMoO$_6$ with
$f=1.06$ is cubic \cite{Borges:99,Ritter:00,Song:01,Kim:01}. We note,
however, that the system is on the brink of a structural phase transition
as a function of $f$. The Ba$_2$FeMoO$_6$ compound is close to a hexagonal
structure. From the collected data listed in Table~\ref{Bigtable} we can
formulate the following empirical rule: For $0.96\lesssim f \lesssim 1.06$
in the majority of cases the double perovskites are cubic/tetragonal. For
$f\lesssim 0.96$, orthorhombic/monoclinic structures are favored, whereas
for $1.06\lesssim f$ a hexagonal structure is preferred. Note, that the
doped manganites of the most intensively investigated composition
La$^{3+}_{2/3}A^{2+}_{1/3}$MnO$_3$ have $f \lesssim 0.95$ depending little
on the chosen coordination number \cite{Hwang:95,Zhou:99}. It is therefore
obvious that structural distortions play a more severe role in the
manganites that are close to a metal-insulator-transition.


\begin{table*}[tbhp]
\begin{widetext}
 \caption{Structure, tolerance factor, and magnetic properties of various double
perovskites. The tolerance $f$ factor was calculated using the SPuDS
simulation software \cite{Lufaso:01}.}
\begin{ruledtabular}
\begin{tabular}{c c c c c c }
material & crystallographic structure & $f$ & magnetic order & $M_{\text{sat}}$(5\,K) & $MR$ @ 50\,kOe  \\
& lattice parameters [{\AA}] & & $T_C$ or T$_N$ [K]  & ($\mu_B/$f.u.)& [\%] \\
\hline \hline
Ca$_2$CrMoO$_6$&orthorhombic, $a=5.49$&0.954 (3+/5+)&$T_C=148$ \cite{Patterson:63} & &\\
& $b=5.36$, $c=7.70$ \cite{Patterson:63}&&&&
 \\ \hline
 Sr$_2$CrMoO$_6$ & $Fm\overline{3}m$, $a=7.84$ \cite{Arulraj:00} & 1.009 (3+/5+)& $T_C=450$
\cite{Arulraj:00} & 0.5 \cite{Moritomo:00} & -5
(40\,K) \cite{Arulraj:00}
 \\ \hline
 Ba$_2$CrMoO$_6$ && 1.070 (3+/5+)& &&
 \\ \hline
 Ba$_3$Cr$_2$MoO$_9$& $P6_3/mmc$, $a=5.69$, $c=1.39$ \cite{Shikano:95}&   & paramagnetic
\cite{Shikano:95}&&
 \\ \hline \hline
Ca$_2$CrReO$_6$ & $P2_1/n$, $a=5.38$, $b=5.46$&0.952 (3+/5+)& $T_C=360$ \cite{Kato:02} & 0.82 \cite{Kato:02} &\\
                   & $c=7.65$, $\beta=90^\circ$ \cite{Kato:02} & & & &   \\ \hline
Sr$_2$CrReO$_6$    & $I4/mmm$, $a=5.52$, $c=7.82$ \cite{Kato:02}& 1.006
(3+/5+)& $T_C=635$ \cite{Kato:02} &  0.86 \cite{Kato:02} &
\\ \hline
Ba$_2$CrReO$_6$ && 1.067 (3+/5+)& &&
 \\ \hline
Ba$_3$Cr$_2$ReO$_9$ & hexagonal, $a=4.94$, $c=13.8$ \cite{Sleight:62}&
  &  &   & \\ \hline\hline
Ca$_2$CrWO$_6$&$P2_1/n$,  $a=5.39$, $b=5.45$& 0.945 (3+/5+) & $T_C=161$&  1.34 & -9  \\
                   & $c=7.66$, $\beta=90.1^\circ$ &&&&\\
\hline Sr$_2$CrWO$_6$ & $Fm\overline{3}m$, $a=7.82$ & 0.99 (3+/5+)&
$T_C=458$ & 1.11 & -48\\ \hline Ba$_2$CrWO$_6$&
$P\overline{6}2c$, $a=5.70$, $c=13.99$&1.059  (3+/5+)& $T_C=145$&  0.02&0\\
\hline Ba$_3$Cr$_2$WO$_9$ & $P\overline{6}2c$, $a=5.69$, $c=13.99$
\cite{Shikano:95}&& paramagnetic \cite{Shikano:95}&&\\ \hline\hline
Sr$_2$MnMoO$_6$&$Fm\overline{3}m$, $a=8.01$ \cite{Moritomo:00}&0.999 (3+/5+)& $T_N=12$ \cite{Itoh:96}&&\\
               &                                              &0.958 (2+/6+)& &&\\ \hline\hline
Sr$_2$MnReO$_6$ & $Fm\overline{3}m$, $a=8.00$ \cite{Popov:03}& 0.997 (3+/5+)& $T_C=120$ \cite{Popov:03} &  & -10 \cite{Popov:03} \\
                &                                               & 0.949 (2+/6+)&            & & (100\,K)   \\ \hline
Ba$_2$MnReO$_6$ & $Fm\overline{3}m$, $a=8.18$ \cite{Popov:03} & 1.057 (3+/5+) & $T_C=120$ \cite{Popov:03} &  & +14 \cite{Popov:03} \\
                &                                               & 1.006 (2+/6+)&            &      & (80\,K) \\ \hline\hline
Ca$_2$MnWO$_6$& $P2_1/n$,  $a=5.46$, $b=5.65$&0.936 (3+/5+)& $T_C=45$&&\\
                   & $c=7.80$, $\beta=90.2^\circ$ \cite{Azad:01b} &  0.904 (2+/6+)& $T_N=16$ \cite{Azad:01b}&&\\ \hline
Sr$_2$MnWO$_6$      & $P4_2/n$,  $a=8.012$, $c=8.01$ \cite{Azad:01c}& 0.990 (3+/5+)& $T_C=40$&&\\
                   &                                       & 0.956 (2+/6+) &  $T_N=13$ \cite{Azad:01c}&&\\ \hline
Ba$_2$MnWO$_6$& $Fm\overline{3}m$, $a=8.20$ \cite{Azad:01a}&1.049 (3+/5+)&$T_C=45$&&\\
&  &  1.014 (2+/6+)& $T_N=10$ \cite{Azad:01a}&&\\ \hline \hline
Ca$_2$FeMoO$_6$     & $P2_1/n$, $a=5.41$, $b=5.52$, & 0.946 (3+/5+)&   $T_C=365$ \cite{Ritter:00} & 3.51 \cite{Ritter:00}& -29 \cite{Dai:01} \\
                   & $c=7.71$, $\beta=90.0^\circ$ \cite{Ritter:00}&&&&\\ \hline
Sr$_2$FeMoO$_6$     & $I4/mmm$, $a=5.58$, $c=7.89$ \cite{Tomioka:00}&1.000 (3+/5+)&$T_C=420$ \cite{Tomioka:00}&3.7 \cite{Balcells:01}& -37 \cite{Sarma:00a}\\
\hline
Ba$_2$FeMoO$_6$&$Fm\overline{3}m$, $a=8.06$~\cite{Borges:99}&1.060 (3+/5+)&$T_C=367$ \cite{Borges:99}&3.53 \cite{Ritter:00}  &  -25~\cite{Park:01}\\
 &&&&& (8\,kOe)\\ \hline \hline
Ca$_2$FeReO$_6$     & $P2_1/n$, $a=5.40$, $b=5.52$           &  0.943 (3+/5+)    &  $T_C=540$ \cite{Westerburg:02}&2.24 \cite{Prellier:00}& 0 \cite{Gopalakrishnan:00} \\
                   & $c=7.68$ $\beta=90.02^\circ$~\cite{Gopalakrishnan:00}&  &                               &                                                    &                    \\
\hline
Sr$_2$FeReO$_6$&$Fm\overline{3}m$, $a=7.89$ \cite{Gopalakrishnan:00}&0.997 (3+/5+)&$T_C=400$ \cite{Kato:02b}&2.7 \cite{Kobayashi:99}& -26 \cite{Gopalakrishnan:00} \\
\hline
Ba$_2$FeReO$_6$     & $Fm\overline{3}m$, $a=8.06$~\cite{Gopalakrishnan:00}&1.057 (3+/5+)&  $T_C=315$ \cite{Prellier:00}& 3.04 \cite{Prellier:00}  & -8 \cite{Gopalakrishnan:00}\\
\hline Ba$_3$Fe$_2$ReO$_9$ & hexagonal, $a=5.03$, $c=14.10$
\cite{Sleight:62}&
  &  &   & \\ \hline\hline
Sr$_2$FeWO$_6$      & $P2_1/n$, $a=5.65$, $b=5.61$&0.969 (2+/6+)&$T_N=40$ \cite{Kobayashi:00}&&   \\
                   & $c=7.94$, $\beta=89.99^\circ$ \cite{Azad:01d}&&&&\\ \hline
Ba$_2$FeWO$_6$&$I4m$, $a=5.75$, $c=8.13$ \cite{Azad:01d}&1.028 (2+/6+)&
$T_N\approx20$ \cite{Azad:01d} & & \\
\end{tabular}
 \label{Bigtable}
\end{ruledtabular}
\end{widetext}
\end{table*}

\subsection{Transport Properties}

\begin{figure}[tb]
\centering{%
\includegraphics [width=0.9\columnwidth,clip=]{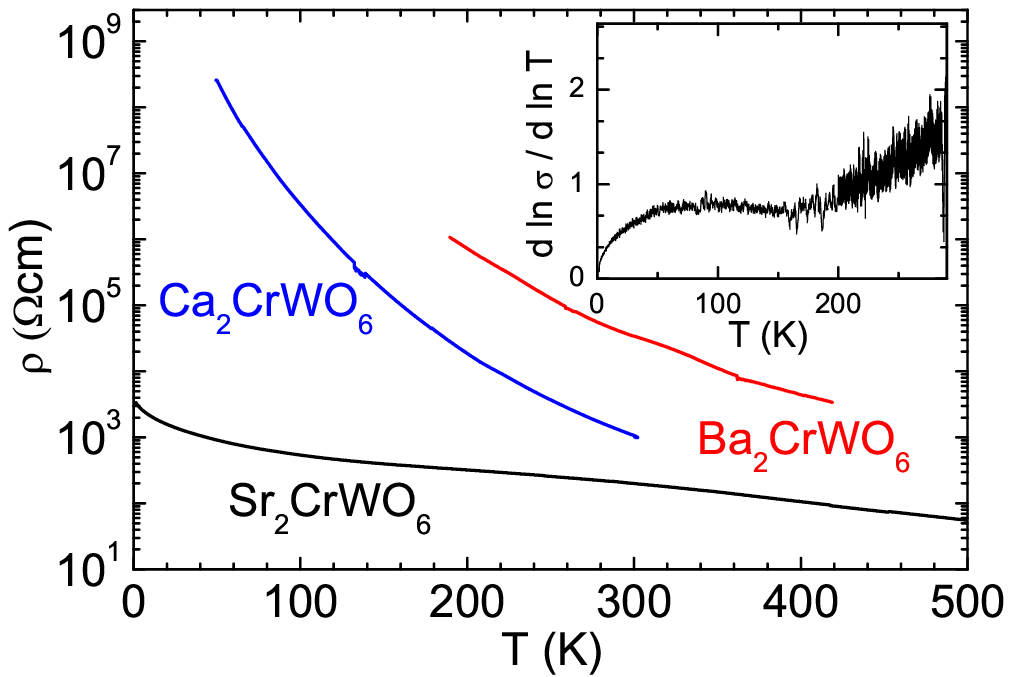}}
 \caption{Resistivity vs temperature for the double perovskites
$A_2$CrWO$_6$ with  $A= \text{Sr, Ba, Ca}$. The inset shows $d\ln \sigma /d
\ln T$ versus $T$ for the Sr$_2$CrWO$_6$ sample. }
 \label{Fig:res}
\end{figure}

Fig.~\ref{Fig:res} shows the temperature dependence of the resistivity for
the double perovskites $A_2$CrWO$_6$ with $A= \text{Sr, Ba, Ca}$. All
samples show an increase of the resistivity with decreasing temperature.
Since the investigated samples are polycrystalline, the influence of grain
boundaries plays an important role. Hence, the observed semiconductor like
resistivity vs temperature curves may be related to the grain boundary
resistance, whereas the intrinsic resistance of the double perovskites may
be metallic. Although the intrinsic resistivity behavior cannot be
unambiguously derived from our measurements, the fact that $d\ln \sigma /d
\ln T \rightarrow 0$ for $T \rightarrow 0$ (see inset of
Fig.~\ref{Fig:res}) provides significant evidence for a metallic behavior
in the Sr$_2$CrWO$_6$ sample. Here, $\sigma$ is the electrical
conductivity. The observed trend that the resistivity increases, if Sr$_2$
is replaced by Ca$_2$ and even more by Ba$_2$, can be understood as
follows: In contrast to the cubic perovskite Sr$_2$CrWO$_6$, Ca$_2$CrWO$_6$
has a distorted perovskite structure with a $B$-O-$B$ bonding angle
deviating from 180$^\circ$. This result in a reduction of the overlap
between the relevant orbitals and, hence, the hopping amplitude. Finally,
the Ba$_2$CrWO$_6$ compound has the highest resistivity most likely due to
its hexagonal structure (see Table~\ref{Table:Samples}).

The magnetotransport properties of Sr$_2$CrWO$_6$ already has been
discussed elsewhere \cite{Philipp:01}. We found, that polycrystalline
samples containing a large number of grain boundaries show a large negative
low-field magnetoresistance, $MR = [R(H) - R(0)]/R(0)$, of up to -41\% at
5\,K. At room temperature, this effect is reduced to a few percent. The
large grain boundary $MR$ effect at low temperatures indicates that
Sr$_2$CrWO$_6$ has a large spin polarization of the charge carries and due
to its high Curie temperature may be an interesting candidate for
magnetoelectronic devices operating room-temperature.

\subsection{Magnetic Properties}

\begin{figure}[tb]
\centering{%
\includegraphics [width=0.9\columnwidth,clip=]{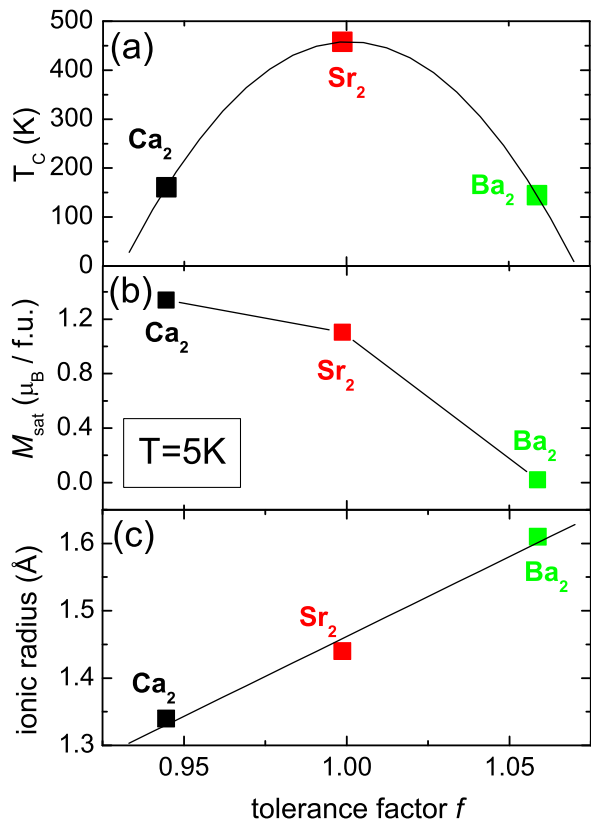}}
 \caption{(a) Curie temperature, (b)
saturation magnetization at 5\,K and an applied field of 7\,T, and (c)
ionic radii for the series $A_2$CrWO$_6$ with  $A = \text{Sr, Ba, Ca}$
plotted versus the tolerance factor $f$.}
 \label{Fig:f}
\end{figure}

In Fig.~\ref{Fig:f} we plot the Curie temperature $T_C$, the saturation
magnetization $M_{\rm sat}$, and the ionic radii versus the tolerance
factor for the series $A_2$CrWO$_6$. It is evident that the Curie
temperature is largest for Sr$_2$CrWO$_6$ ($T_C=458$\,K), whereas it is
suppressed strongly for Ca$_2$CrWO$_6$ ($T_C=161$\,K). We attribute this
fact to the small ionic radius of Ca$^{2+}$, which results in $f \le 1$
and, in turn, in a distorted perovskite structure. This results in a
reduction of the effective hopping interactions between Cr~$3d$ and W~$5d$
states, leading to a reduced spin splitting of the conduction
band~\cite{Sarma:00}. This naturally reduces the magnetic coupling strength
and hence the $T_C$.

The saturation magnetization $M_{\rm sat}$ is known to depend strongly on
the amount $\delta $ of antisites, with $\delta =0$ for no antisites and
$\delta =0.5$ for 50\% antisites or complete disorder. By simply assuming
the presence of antiferromagnetically coupled Cr and W sublattices, a
maximum saturation magnetization of $2\mu_B/\text{f.u.}$ is expected for
$\delta =0$, which is decreasing to zero for $\delta =0.5$. That is, this
assumption leads to \cite{Balcells:01}
\begin{equation}
M_{\rm sat}(\delta ) = (1-2\delta ) \cdot m(\text{Cr$^{3+}$}) - (1-2\delta
) \cdot m(\text{W$^{5+}$}) \;\; ,
   \label{eq:Msat}
\end{equation}
where $M_{\rm sat}$ is the saturation magnetization in units of
$\mu_B/\text{f.u.}$; $m(\text{Cr$^{3+}$})$  and $m(\text{W$^{5+}$})$ are
the magnetic moments of the Cr$^{3+}$ and W$^{5+}$ ions in units of
$\mu_B$, respectively.  Comparing the experimental data to this simple
model prediction leads to a surprisingly good agreement for Sr$_2$CrWO$_6$.
The measured value of $M_{\rm sat} =  1.11\,\mu_B/$f.u. is very close to
the value of $1.08\,\mu_B/$f.u. expected from eq.(\ref{eq:Msat}) for
$\delta =0.23$. Fig.~\ref{Fig:f}b also shows that the saturation
magnetization of Ca$_2$CrWO$_6$ is larger than for Sr$_2$CrWO$_6$ despite
the much lower Curie temperature of the latter. The obvious reason for that
is the lower amount of antisites in Ca$_2$CrWO$_6$. From the measured value
of $\delta =0.13$ we expect $M_{\rm sat} =  1.48\,\mu_B/$f.u. for
Ca$_2$CrWO$_6$ which is slightly larger than the measured value of  $M_{\rm
sat} = 1.34\,\mu_B/$f.u.. The reason for the observation that the
experimental $M_{\rm sat} $ value is below the one predicted by
eq.(\ref{eq:Msat}) most likely is the distorted crystal structure of
Ca$_2$CrWO$_6$.

The magnetic properties of the Ba$_2$CrWO$_6$ compound are completely
different from those of Sr$_2$CrWO$_6$ and Ca$_2$CrWO$_6$. Here, the large
ionic radius of Ba$^{2+}$ enlarges $f$ well above unity. This causes a
structural phase transition towards a hexagonal structure, where the
ferromagnetic interaction is strongly suppressed. Therefore, not only $T_C$
is strongly suppressed ($T_C= 145$\,K, in contrast to $T_C= 458$\,K for
Sr$_2$CrWO$_6$), but also the saturation magnetization ($\sim
0.02\,\mu_B/$f.u.) is close to zero. This clearly indicates the strong
effect of the structural phase transition on the magnetic interaction. We
note that we can exclude the possibility that the very small $M_{\rm sat}$
value is due to strong disorder. Evidently, there should be almost complete
disorder to suppress $M_{\rm sat}$ close to zero. However, in the
Ba$_2$CrWO$_6$ compound the two different $B$ sites establish a certain
amount of order, since almost all W ions occupy the $B_2$ site. Whether the
observed behavior is related to a canted antiferromagnetic phase or simply
to the presence of minority phases has to be clarified.

\begin{figure}[tb]
\centering{%
\includegraphics [width=0.9\columnwidth,clip=]{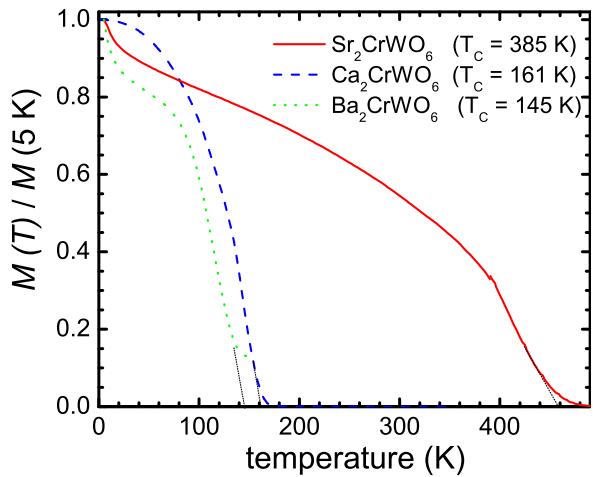}}
 \caption{Normalized magnetization versus temperature for
the series $A_2$CrWO$_6$ with  $A = \text{Sr, Ba, Ca}$. The measurements
were made field cooled in an applied field of $H=100$\,Oe. The black lines
represent are extrapolations of the measured data for the determination of
$T_C$. Due to the small $M_{\rm sat}$ of the Ba$_2$CrWO$_6$ compound, the
zero line is shifted due to a small residual paramagnetic moment in the
applied field.}
 \label{Fig:acwomag}
\end{figure}

Fig.~\ref{Fig:acwomag} shows that the magnetic interactions for all three
compounds $A_2$CrWO$_6$ is ferromagnetic. However, the strongly reduced
saturation magnetization of Ba$_2$CrWO$_6$ and the upturn in magnetization
vs. temperature curve at low temperatures indicates that ferromagnetic
interactions are small and that there may be paramagnetic regions in the
sample. The substitution of Sr by Ca evidently results in a strong
reduction of $T_C$, however, the Ca$_2$CrWO$_6$ samples are still clearly
ferromagnetic. Again, the important point is that only the compound with
$f\simeq 1$ has optimum magnetic properties with respect to applications in
magnetoelectronics.

We also have performed measurements of the coercive field $H_c$. At low
temperature we obtained $H_c \simeq 450$\,Oe for the Sr$_2$CrWO$_6$
compound and $H_c \simeq 6000$\,Oe for the Ca$_2$CrWO$_6$ compound.
Furthermore, for the Sr$_2$CrWO$_6$ the coercive field was found to depend
on the preparation conditions. For example, lowering the final firing
temperature from 1550$^\circ$C to 1300$^\circ$C was found to increase $H_c$
from 450 to 1200\,Oe.  In agreement with the findings for the doped
manganites \cite{Zhou:99}, the coercive field increases with decreasing
$f$. This observation can be easily understood, since the induced
structural distortions can effectively act as pinning centers for domain
wall movement. In those cases where a large remanent magnetization and/or
coercive field are important for applications, the use of double
perovskites with reduced values of $f$ obtained by suitable substitution on
the $A$-site may be desirable.

Comparing our results for the series $A_2$CrWO$_6$ with  $A = \text{Sr, Ba,
Ca}$ to other double perovskite compounds summarized in
Table~\ref{Bigtable}, it is evident that the suppression of $T_C$ as a
function of the deviation of the tolerance factor from its ideal value of
$f=1$ is a general trend: It is only weak for the series $A_2$FeMoO$_6$
with  $A = \text{Sr, Ba, Ca}$, where $T_C$ varies between 310\,K and 420\,K
\cite{Galasso:69,Borges:99,Ritter:00,Song:01,Kim:01}. However, it is also
strong for the series $A_2$CrReO$_6$ with $T_C= 635$\,K for Sr$_2$CrReO$_6$
and $T_C= 360$\,K for Ca$_2$CrReO$_6$ \cite{Kato:02}. In general, a high
Curie temperature can only be realized in double perovskites of the
composition $A_2BB^\prime$O$_6$ having a tolerance factor close to unity.
This is realized in the different systems for $A_2 = \text{Sr}_2$. For $f$
well below unity, the Curie temperature is drastically reduced in agreement
with what is found for the doped manganites \cite{Hwang:95}. For the double
perovskites, the system (Sr$_{1-y}$Ca$_y$)$_2$FeReO$_6$ is an exception of
the general rule \cite{Kato:02b}: Here, the Ca$_2$FeReO$_6$ compound has
the highest $T_C$, although the tolerance factor decreases continuously
from $f=0.997$ for Sr$_2$FeReO$_6$ to $f=0.943$ for Ca$_2$FeReO$_6$ on
substituting Sr by Ca. We note, however, that Ca$_2$FeReO$_6$ is a unique
material as it is a ferromagnetic insulator and that there may be an other
mechanism causing the high ordering temperature \cite{Prellier:00}.

\begin{figure}[tb]
\centering{%
\includegraphics [width=0.9\columnwidth,clip=]{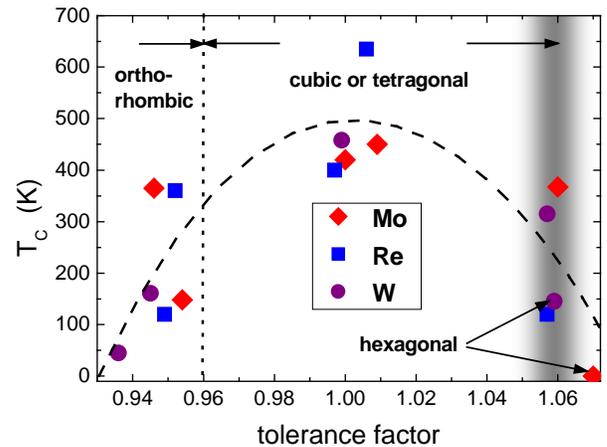}}
 \caption{Curie temperature of the different double perovskite materials
listed in Table~\ref{Bigtable} plotted versus their tolerance factor.
Re-compounds: squares, W-compounds: circles, Mo-compounds: diamonds. The
broken line only serves as a guide to the eyes.}
 \label{Fig:tctoleranceall}
\end{figure}

In Fig.~\ref{Fig:tctoleranceall} we have plotted the Curie temperatures of
the double perovskite materials listed in Table~\ref{Bigtable} versus their
tolerance factor omitting the ferromagnetic insulator Ca$_2$FeReO$_6$.
Despite the significant spread of data that may be partially caused be
different sample quality, it is evident that a maximum Curie temperature is
obtained for the systems having a tolerance factor of $f \simeq 1$. These
systems have a cubic/tetragonal symmetry. As shown in
Fig.~\ref{Fig:tctoleranceall}, for $f<1$ there is a transition to
orthorhombic structures for $f \sim 0.96$, whereas for $f>1$ there is a
transition to a hexagonal structure at $f \sim 1.06$. However, according to
data from literature this transition seems to be smeared out as indicated
by the shaded area.

\subsection{Underlying Physics}

\begin{figure}[tb]
\centering{%
\includegraphics [width=0.9\columnwidth,clip=]{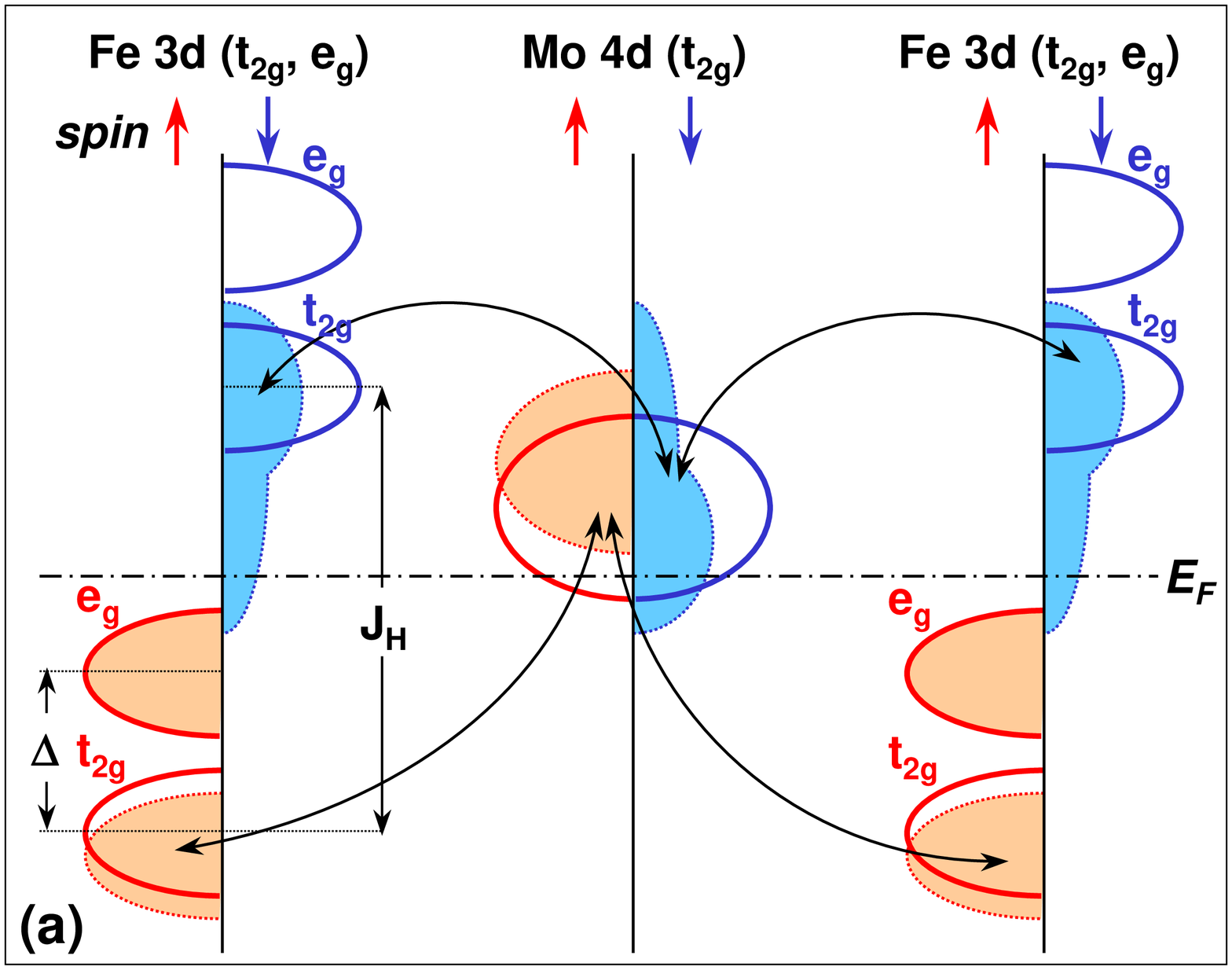}
 \\
\includegraphics [width=0.9\columnwidth,clip=]{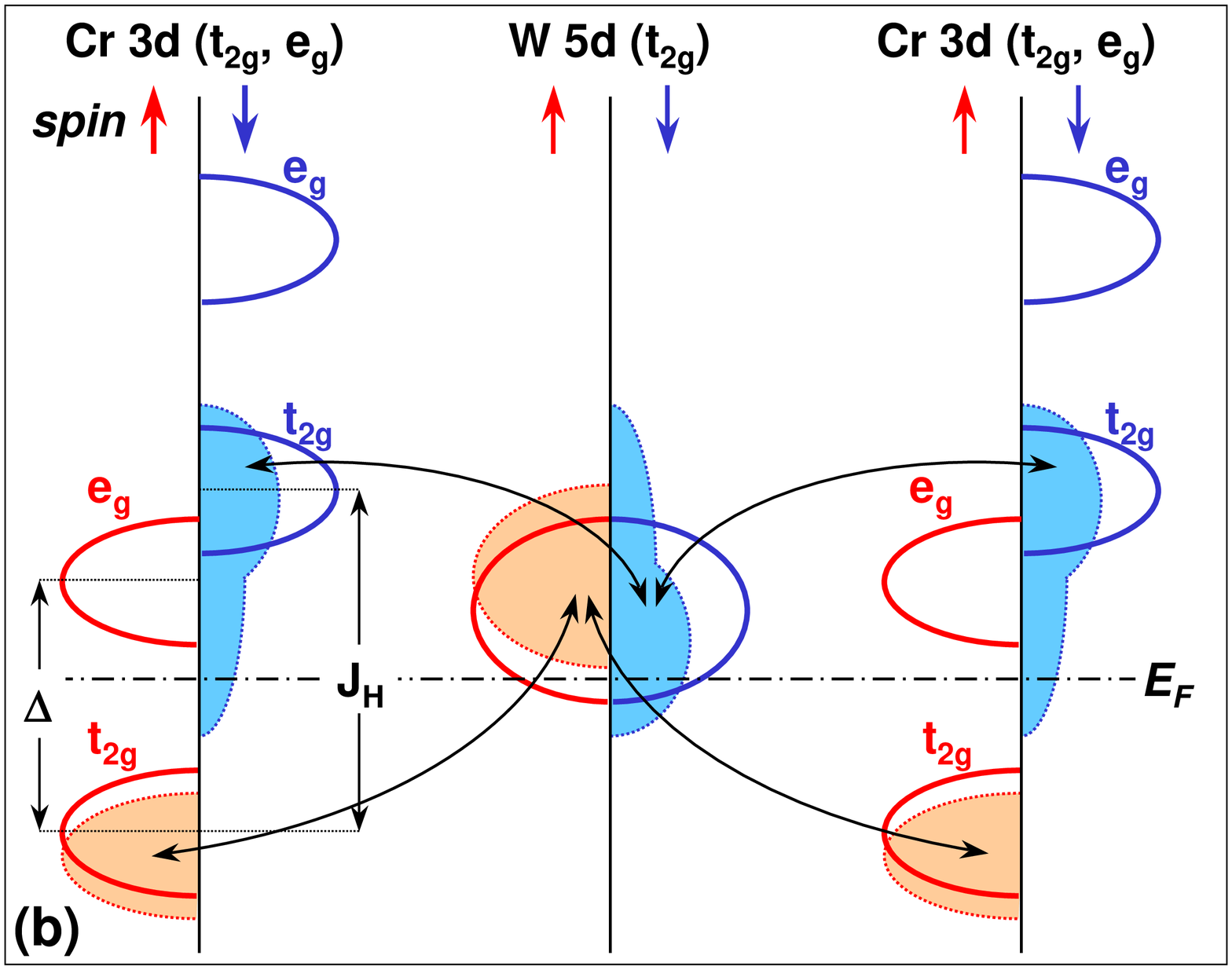}}
 \caption{Sketch of the band structure for the illustration of
the mechanism stabilizing the ferromagnetic state in the double
perovskites. (a) shows the situation for $A_2$FeMoO$_6$. Here, the Fermi
energy lies between the exchange split $3d$ spin-up and -down states.  In
(b) we show the situation for $A_2$CrWO$_6$. Here, the Fermi level lies in
the crystal-field split Cr~$3d$ spin-up band between the $t_{2g}$ and the
$e_g$ levels. The solid lines mark the bands without hybridization, whereas
the shaded areas denote the bands with hybridization. The arrows connect
the hybridizing bands. The hybridization between the $e_g$ level has been
neglected. }
 \label{Fig:models}
\end{figure}

We briefly summarize the measured structural, transport and magnetic
properties and discuss the underlying physics. Recently, Sarma {\it et al.}
proposed an interesting model explaining the origin of the strong
antiferromagnetic coupling between Fe and Mo in Sr$_2$FeMoO$_6$ in terms of
an strong effective exchange enhancement at the Mo site due to a
Fe~$3d$-Mo~$4d$ hybridization \cite{Sarma:00}. Kanamori and Terakura
extended this idea to explain ferromagnetism in many other systems, where
nonmagnetic elements positioned between high-spin $3d$ elements contribute
to the stabilization of ferromagnetic coupling between the $3d$ elements
\cite{Kanamori:01}. The essence of this model is summarized in
Fig.~\ref{Fig:models}a for the case of Sr$_2$FeMoO$_6$. Without any hopping
interactions, the Fe$^{3+}$ $3d^5$ configuration has a large exchange
splitting of the $3d$ level in the spin-up and spin-down states and there
is also a crystal field splitting $\Delta$ into the $t_{2g}$ and the $e_g$
states (see Fig.~\ref{Fig:models}a). The exchange splitting of the
Mo$^{5+}$~$4d^1$ configuration (better the Mo-$4d$-O-$2p$ hybridized
states) is vanishingly small, however, there is a large crystal field
splitting (the $e_g$ states are several eV above the $t_{2g}$ states and
not shown in Fig.~\ref{Fig:models}a). The interesting physics occurs on
switching on hopping interactions, which result in a finite coupling
between states of the same symmetry and spin. The hopping interaction not
only leads to an admixture of the Fe~$3d$ to the Mo~$4d$ states, but more
importantly to a shift of the bare energy levels. As shown in
Fig.~\ref{Fig:models}a, the delocalized Mo~$t_{2g}$ spin-up states are
pushed up, whereas the Mo~$t_{2g}$ spin-down states are pushed down. This
causes a finite spin polarization at the Fermi level (actually 100\% in
Fig.~\ref{Fig:models}a) resulting from the hopping interactions. This
kinetic energy driven mechanism leads to an antiferromagnetic coupling
between the delocalized Mo~$4d$ and the localized Fe~$3d$ electrons, since
the energy is lowered by populating the Mo~$4d$ spin-down
band~\cite{Sarma:00}. The magnitude of the spin polarization derived from
this mechanism obviously is governed by the hopping strength and the charge
transfer energy between the localized and the delocalized
states~\cite{Kanamori:01}.

The question now arises, whether the above model also applies for the
$A_2$CrWO$_6$ compounds. The key concept of the model is the energy gain
contributed by the spin polarization of the nonmagnetic element (now W)
induced by the hybridization with the magnetic transition metal (now Cr).
It has been pointed out by Sarma~\cite{Sarma:00} that the underlying
mechanism will always be operative, whenever the conduction band is placed
within the energy gap formed by the large exchange splitting of the
localized electrons at the transition metal site. As shown in
Fig.~\ref{Fig:models}b, this is also the case for the $A_2$CrWO$_6$
compounds: the W~$5d$ band resides in between the exchange split Cr~$3d$
($t_{2g}$) spin-up and spin-down bands. The schematic band structure of
Fig.~\ref{Fig:models}b has been confirmed by band structure calculations
presented below. The only difference between the system $A_2$CrWO$_6$ and
the systems $A_2$FeMoO$_6$ or $A_2$FeReO$_6$ is the fact that in the former
the majority spin band is only partially full. For Cr$^{3+}$, ($3d^3,
S=3/2$) only the Cr $t_{2g}$ levels are occupied. In contrast, for
Fe$^{3+}$ ($3d^5, S=5/2$) both the Fe $t_{2g}$ and Fe $e_{g}$ levels are
occupied, that is, the majority spin band is completely filled. Band
structure calculations show that the crystal field splitting in the Cr
compounds ($\sim 2$\,eV) is slightly larger than in the Fe
compounds~\cite{Jeng:03}. On the other hand, the exchange splitting in the
Cr~$3d$ bands is somewhat smaller than for the Fe~$3d$ bands due to the
valence configuration Cr~$3d^3$ with less electrons and weaker Hund's
coupling. Taking these facts into account we have to split up the Cr~$3d$
spin-up and spin-down band into two separate $3d$ ($t_{2g}$) and $3d$
($e_{g}$) bands with the Fermi level lying in the gap between the bands as
shown in Fig.~\ref{Fig:models}b. Indeed band structure calculations (see
below) show that the Cr~$3d$ ($e_g$) spin-up band is about 0.5\,eV above
the Fermi level. However, the above mechanism still works as long as the
W~$5d$ ($t_{2g}$) band is placed within the energy gap between the Cr~$3d$
($t_{2g}$) spin-up and the Cr~$3d$ ($t_{2g}$) spin-down band. Then, again
the W~$5d$ ($t_{2g}$) levels would hybridize with the Cr~$3d$ ($t_{2g}$)
levels resulting in a negative spin polarization of the nonmagnetic element
W~$5d$ ($t_{2g}$) and a stabilization of ferromagnetism and half-metallic
behavior. We note that no hybridization takes place between the Cr~$3d$
($e_g$) spin-up band and the W~$5d$ ($t_{2g}$) spin-up band due to the
different symmetry of these levels. Therefore, the exact position of the
Cr~$3d$ ($e_g$) spin-up band is not relevant. We also note that due to the
large crystal field splitting for the W~$5d$ band, the W~$3d$ ($e_g$) band
is several eV above the W~$3d$ ($t_{2g}$) band and not shown in
Fig.~\ref{Fig:models}b. Summarizing our discussion we can state that the
essential physical mechanism leading to ferromagnetism is very similar for
the $A_2$CrWO$_6$ and the $A_2$FeMoO$_6$ compounds. Comparing
Fig.~\ref{Fig:models}a and b we see that the only difference is an upward
shift of the $3d$ bands for the $A_2$CrWO$_6$ compounds.

We now discuss the experimental results in context with the models
discussed above.  We first discuss the strong dependence of $T_C$ on the
tolerance factor, which in turn is intimately related to the bond angles.
It is well known that in perovskite type transition metal oxides in general
the increase of the $B$-O bond length and the deviation of the $B$-O-$B$
bond angle from $180^\circ$ has the effect of a reducing the hybridization
matrix element $t$. This is caused by the reduction of the overlap of the
oxygen 2$p$ and transition-metal $d$ states \cite{Imada:98}, which in turn
results in a reduction of the bare electron bandwidth $W$. Since the
weakening of the hybridization causes a reduction of the energy gain
stabilizing ferromagnetism, we expect a significant decrease of $T_C$ with
increasing deviation of the tolerance factor from $f=1$, or equivalently
with an increasing deviation of the $B$-O-$B$ bond angle from $180^\circ$.
This is in good qualitative agreement with our results on the $A_2$CrWO$_6$
compound (see Fig.~\ref{Fig:f}a) and the collected data plotted in
Fig.~\ref{Fig:tctoleranceall}. Theoretical models providing a quantitative
explanation have still to be developed.

It is known that for a $180^\circ$ bond angle, $t$ decreases with
increasing bond length $d_{\rm B-O}$ roughly as $t \propto 1/d_{\rm
B-O}^{3.5}$. Therefore, the effective hopping amplitude between the
transition metal ions $B$ has an even stronger dependence on $d_{\rm B-O}$.
Hence, a significant increase of $t$ and, in turn, $T_C$ is expected with a
reduction of the bond length. Applying this consideration to the
investigated series $A_2$CrWO$_6$ we would expect the largest $T_C$ for the
Ca$_2$CrWO$_6$ compound due to its smallest cell volume and, hence,
shortest bond length. However, experimentally the largest $T_C$ was found
for Sr$_2$CrWO$_6$. This is caused by the fact that $t$ not only depends on
the bond length but also strongly on the bond angle. For the cubic double
perovskite Sr$_2$CrWO$_6$ with $f\simeq 1$ the bond angle has the ideal
value of $180^\circ$, whereas for the distorted double perovskite
Ca$_2$CrWO$_6$ with $f<1$ the bond angle significantly deviates from this
ideal value. For the double perovskites lattice effects come into play also
for large tolerance factors $f \gtrsim 1.06$, where in the hexagonal
structure the formation of dimers suppresses strongly ferromagnetism and
enhances antiferromagnetic interactions. Furthermore, the magnetic
interactions are also weakened due to the increased distance (bond length)
between the magnetic ions. We finally would like to mention that in a
recent work on $A_2$FeMoO$_6$, a linear correlation, $T_C \propto W$,
between $T_C$ and the bare one-electron bandwidth $W \propto t$ has been
found by Ritter {\em et al.}~\cite{Ritter:00}.

Summarizing our discussion we can state that our findings for the series
$A_2$CrWO$_6$ can be extended to the double perovskites in general. The
data summarized in Table~\ref{Bigtable} and Fig.~\ref{Fig:tctoleranceall}
clearly indicated that for most double perovskites a maximum $T_C$ is
obtained for a tolerance factor of $f \simeq 1$ corresponding to an about
cubic perovskite structure with a bond angle close to $180^\circ$. This
optimum situation in most cases is realized in the Sr$_2BB^\prime$O$_6$
compounds. The requirement $f\simeq 1$ for optimum $T_C$ in the double
perovskite is different for the doped manganites. Here, a maximum $T_C$ is
achieved for compounds with $f\lesssim 0.95$, that is, for a significantly
distorted perovskite structure.  Hwang {\em et al.}~\cite{Hwang:95} have
shown that the highest $T_C$ in doped manganites is obtained for $f\simeq
0.93$ (in a more precise analysis by Zhou {\em et al.}~\cite{Zhou:99}
slightly larger values for $f$ have been derived based on a coordination
number of 9). For a tolerance factor below the optimum value, a dramatic
decrease of $T_C$ has been found. Whether or not the differences between
the manganites and the double perovskites are related to the different
mechanism stabilizing ferromagnetism in these compounds has to be
clarified. We note, however, that the different behavior of the manganites
and the double perovskites is likely to be related to different mechanisms.
In manganites the polarization of the conduction band is driven by the
Hund's coupling (intra atomic) between the $t_{2g}$ and the $e_g$ states,
which is not sensitive to the details of the band structure. In contrast,
in the double perovskites the spin polarization of the conduction band is
itself dependent on the band structure and therefore is more intimately
linked to it.

\section{Electron doping in $\bf Sr_2CrWO_6$}

The effect of carrier (electron) doping in Sr$_2$CrWO$_6$ was studied in a
series of Sr$_{2-x}$La$_x$CrWO$_6$ samples with $x=0$, $0.1$, $0.3$ and
$0.5$. In our experiments trivalent La$^{3+}$ is chosen to replace the
Sr$^{2+}$ ions because the ionic radius of La$^{3+}$ ($r_{\rm La^{3+}} =
1.36$\,{\AA}) is similar to that of Sr$^{2+}$ ($r_{\rm Sr^{2+}} = 1.44$\,{\AA}). As
a result, there are only a small variations of the lattice parameters and
the tolerance factor $f$ on changing the doping level from $x=0$ to
$x=0.5$. X-ray analysis showed that the lattice parameter slightly
decreases from $7.818$\,{\AA}\ ($x=0$) to $7.804$\,{\AA}\ ($x=0.5$) as expected
since the ionic radius of La$^{3+}$ is smaller than that of Sr$^{2+}$.
However, the crystal structure of all samples remained cubic. This means
that the structural changes are small on varying the doping level. In this
way the effect of doping can be studied without being strongly influenced
by structural effects.

\begin{figure}[tb]
\centering{%
\includegraphics [width=0.9\columnwidth,clip=]{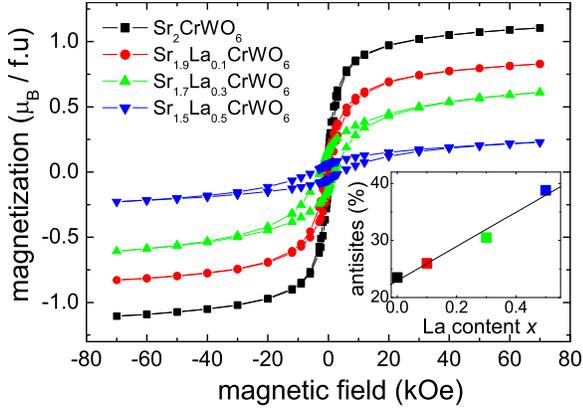}}
 \caption{Magnetization vs. applied magnetic field at $T=5$\,K for the series
Sr$_{2-x}$La$_x$CrWO$_6$ with $x=0$, $0.1$, $0.3$ and $0.5$.  The inset
shows the amount of antisites determined a Rietveld refinement of the x-ray
data.}
 \label{Fig:La}
\end{figure}

In Fig.~\ref{Fig:La} the magnetization curves of the series
Sr$_{2-x}$La$_x$CrWO$_6$ are shown for  $T=5$\,K. All samples of this
series have been fired between 1530 and $1550^\circ$C in reducing
atmosphere. For the undoped sample this results in a smaller Curie
temperature of about 390\,K as compared to about 460\,K for the samples
fired at $1300^\circ$C. The reason for that most likely is a slightly
smaller oxygen content in the sample fired at a higher temperature. Since a
reduced oxygen content corresponds to an effective electron doping, for the
doping series we compare only samples prepared under identical conditions.
We find that the saturation magnetization $M_{\rm sat}$ decreases from
$1.11\,\mu_B$/f.u. for $x=0$ to $0.23\,\mu_B$/f.u. for $x=0.5$.  As shown
in the inset of Fig.~\ref{Fig:La}, with increasing doping level also the
amount of antisites is increasing. At present, we only have a plausible
explanation for the observed increase of the amount of antisites with
increasing doping level, which so far could not be unambiguously proven.
However, since La doping is expected to result in a reduction of the
differences in the valence states of Cr and W, it evidently results in a
reduction of the differences in the ionic radii of Cr and W ($r_{\rm
Cr^{2+}}/ r_{\rm W^{6+}} =0.73/0.63$ and $r_{\rm Cr^{3+}}/ r_{\rm W^{5+}}
=0.615/0.62$). Therefore, increasing the doping level results in more
similar ionic radii of Cr and W paving the way for the creation of Cr/W
antisites. Since the substitution of Sr$^{2+}$ by La$^{3+}$ results both in
electron doping and an increase of antisites, it is not possible to
unambiguously attribute the measured decrease in $M_{\rm sat}$ to either
the increasing doping level or the increase of antisites alone. As will be
discussed in the following, for the series Sr$_{2-x}$La$_x$CrWO$_6$ the
measured reduction of $M_{\rm sat}$ most likely is caused by both electron
doping and disorder.

We first discuss the expected reduction of $M_{\rm sat}$ due to the
increasing amount of antisites. We note that that several authors have
found a reduction of $M_{\rm sat}$ following the increase of antisites
\cite{Navarro:01,Balcells:01}. This behavior is consistent with simple
Monte Carlo simulation studies \cite{Ogale:99}. However, also more
complicated models \cite{Sarma:00,Kanamori:01,Fang:01} predict a reduction
of $M_{\rm sat}$ with increasing amount of antisites due to charge transfer
effects \cite{Saha:01}. In a first approach we can analyze our data using
eq.(\ref{eq:Msat}). With the measured $\delta$ values for the amount of
antisites we can calculate $M_{\rm sat}(\delta )$. We find that the
calculated $M_{\rm sat}$ values are significantly larger than the measured
ones. This suggests that the observed reduction of the saturation
magnetization cannot be explained by the increasing amount of antisites
alone (at least within the simple model yielding eq.(\ref{eq:Msat})).
However, we also have to keep in mind that electron doping itself
contributes to the reduction of $M_{\rm sat}$. According to the
illustration given in Fig.~\ref{Fig:models}a it is evident that electron
doping in the $A_2$FeMoO$_6$ system increases the spin-down magnetic moment
at the Mo site, indicating that the electrons are filled into the Mo
$4d_{\downarrow}$ band and thereby reduce $M_{\rm sat}$
\cite{Moritomo:00e}. According to Fig.~\ref{Fig:models}b, the same
mechanism holds for the Sr$_{2-x}$La$_x$CrWO$_6$ system. That is, our
results suggest that the observed reduction of the saturation magnetization
with increasing La$^{3+}$ substitution is caused both by an increase of the
amount of antisites and an increase of the number of conduction electrons.

Summarizing our discussion of the saturation magnetization we would like to
emphasize that a variation of the doping level in most cases is correlated
with a variation of the amount of disorder, since doping contributes to a
reduction of the difference of the valence states of Cr and W (Fe and Mo)
what in turn results in an increasing amount of antisites
\cite{Anderson:93}. Unfortunately, due to this fact the experimental
situation is not completely clear. While in \cite{Navarro:01} for
Sr$_2$FeMoO$_6$ and also in our study for Sr$_2$CrWO$_6$ La doping is
clearly correlated with higher disorder, in \cite{Moritomo:00e} the amount
of antisites seems to be constant in Sr$_{2-x}$La$_x$FeMoO$_6$ for almost
the whole doping series from $x=0$ to $x=0.3$. For the system
Sr$_2$CrWO$_6$, the suppression of $M_{\rm sat}$ is stronger than for the
system Sr$_2$FeMoO$_6$, probably due to the fact that the Cr and W ions are
easier to disorder because of their similar ionic radii. We finally note
that in general both increasing doping and disorder can destroy the
underlying half-metallic ferromagnetic state in double perovskites, thereby
significantly populating/depopulating the different spin channels leading
to a sharp decrease in $M_{\rm sat}$ \cite{Saha:01}.

\begin{figure}[tb]
\centering{%
\includegraphics [width=0.9\columnwidth,clip=]{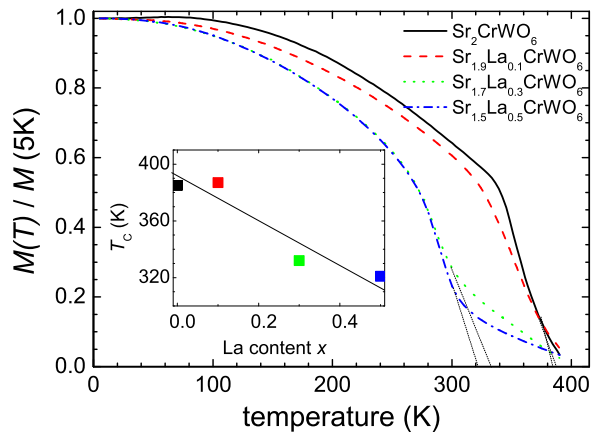}}
 \caption{Normalized magnetization vs temperature for
the series Sr$_{2-x}$La$_x$CrWO$_6$ with $x=0$, $0.1$, $0.3$, and $0.5$.
The measurements were made field cooled in $H=100$\,Oe. The inset shows the
Curie temperature as a function of the La doping. }
 \label{Fig:La-Tc}
\end{figure}

We next discuss the influence of La doping on the Curie temperature. It is
evident that an enhancement of $T_C$ by doping would be of great importance
for possible applications of the double perovskites in spintronic devices.
However, both the experimental data on the variation of $T_C$ with electron
doping as well as the theoretical interpretation is controversial at
present. In single crystals of Sr$_{2-x}$La$_x$FeMoO$_6$ Moritomo {\em et
al.}~have found that $T_C$ does not change as a function of La doping
$x\leq 0.3$ \cite{Moritomo:00e}. In contrast, Navarro {\em et al.}~have
reported a considerable increase of $T_C$ of about 70\,K in ceramic samples
of Sr$_{2-x}$La$_x$FeMoO$_6$ investigating a wider doping range $0\leq
x\leq1$ \cite{Navarro:01}. In our study on the system
Sr$_{2-x}$La$_x$CrWO$_6$ we have found a reduction of $T_C$ of about 80\,K
in the doping range $0\leq x\leq0.5$ as can be seen in
Fig.~\ref{Fig:La-Tc}. We note that part of the experimental discrepancies
may be related to large error bars in the determination of $T_C$. In
particular, great care has to be taken over the determination of $T_C$,
since the transition from zero magnetization to finite magnetization is
smeared out considerably due to finite applied magnetic fields and
parasitic phases. It is evident from Fig.~\ref{Fig:La-Tc} that in all our
samples there are minor phases with optimum $T_C$ close to about 400\,K.
However, it is also evident that La doping reduces $T_C$ of the major part
of the sample. That is, different experimental results on $T_C$ may be in
part related to different ways of measuring and analyzing the data.

From the theoretical point of view one expects both an increase and
decrease of $T_C$ with increasing doping. On the one hand, within the model
presented in \cite{Sarma:00,Kanamori:01,Fang:01} $T_C$ is expected to be
rather reduced than enhanced by electron doping due to the fact that the
possible energy gain by shifting electrons from spin-up band into the
spin-down band is reduced \cite{Phillips:03}. This, in turn, reduces the
stability of the ferromagnetic phase in agreement with our results.
However, the role of the increasing amount of antisites with increasing
doping still has to be clarified. On the other hand, in a double exchange
model the increase of the number of conduction electrons promoted by La
doping is expected to enhance the double exchange interaction leading to an
increase of $T_C$. Within this model the ferromagnetic interaction arises
from the double exchange interaction between the localized moments on
Cr$^{3+}$ sites ($3d^3$, $S=3/2$) mediated by itinerant electron provided
by the W$^{5+}$ ions ($5d^1$, $S=1/2$). According to the band structure
calculation presented below, the Cr $t_{2g}$ spin-up subband is completely
filled and it is the electron in the $t_{2g}$ spin-down subband of Cr and W
which mediates the double exchange interaction. In general, an increasing
number of electrons in the spin-down subband is expected to strengthen the
double exchange interaction leading to an increase of $T_C$ in conflict
with our experimental findings. That is, the observed decrease of $T_C$
with increasing doping level seems to support the model presented in
\cite{Sarma:00,Kanamori:01,Fang:01} (see also Fig.~\ref{Fig:models}).
However, we have to take into account that by La doping besides the number
of electrons we also increase the amount of disorder. The latter may weaken
the double exchange interaction sufficiently to result in an effective
decrease of $T_C$ also within a double exchange based model. Further work
is required to clarify this in more detail.

\section{Band Structure Calculations and Optical Measurements on $\bf Sr_2CrWO_6$}
\subsection{Band Structure Calculations}

The half-metallicity of ferromagnetic materials is one of the important
ingredients required for applications. In order to obtain a conclusive
picture regarding the spin polarization at the Fermi level it is important
to compare band structure calculations with experimental results. Here, we
compare the results of band structure calculations based on {\em ab initio}
methods to experimental results on the optical reflectivity and
transmissivity.

\begin{figure}[tb]
\centering{%
\includegraphics [width=0.9\columnwidth,clip=]{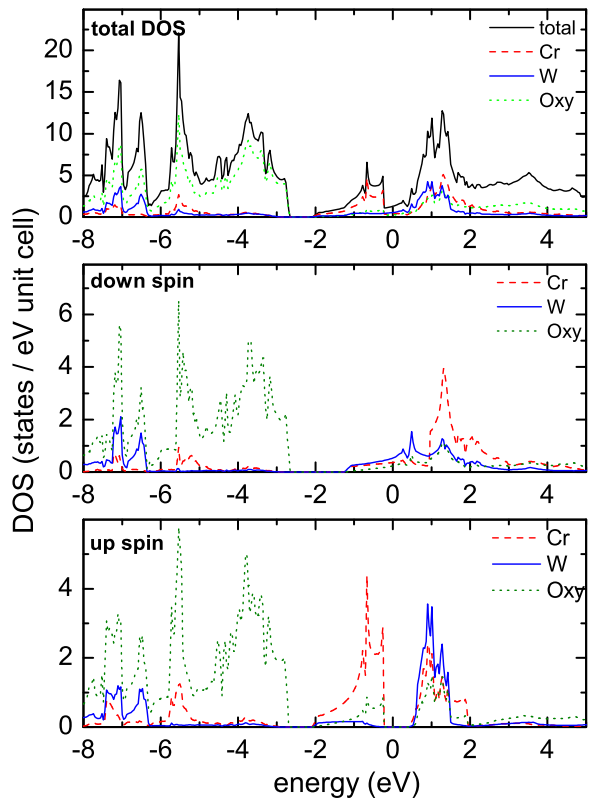}}
 \caption{Density of states obtained from {\em ab initio} band structure
calculations for Sr$_2$CrWO$_6$ plotted as a function of energy with
$E_{\rm Fermi}=0$.}
 \label{Fig:bandstructure}
\end{figure}

In Fig.~\ref{Fig:bandstructure} the density of states of Sr$_2$CrWO$_6$ is
plotted versus energy. The band structure has been
calculated~\cite{Sarma:00,Saha:01} within the linear muffin-tin orbital
(LMTO) method using the atomic sphere approximation (ASA). A detailed
discussion of this method can be found elsewhere
\cite{Anderson:84,Anderson:85}. As a key result of the calculation we
obtain a gap of about $0.7$\,eV in the spin-up band around the Fermi level
(compare also Fig.~\ref{Fig:models}b). This gap corresponds to gap between
the crystal field split Cr~$3d$ ($t_{2g}$) and Cr~$3d$ ($e_{g}$) subband.
On the other hand, the spin-down states are available at the Fermi-level.
The broad band corresponds to the hybridized Cr~$3d$ ($t_{2g}$) and W~$5d$
($t_{2g}$) levels.  That is, according to the band structure calculation we
expect a half-metallic behavior for the double perovskite Sr$_2$CrWO$_6$. A
similar result but with slightly larger band gaps has been reported
recently by Jeng {\em et al.} \cite{Jeng:03}. The result of the band
structure calculation is in agreement with the model assumptions made
above. The Cr~$3d$ ($t_{2g}$) spin-up subband is completely filled and
there is a gap to the Cr~$3d$ ($e_{g}$) spin-up subband lying about 0.5\,eV
above the Fermi level.  The W~$5d$ ($t_{2g}$) spin-up band also is mainly
above the Fermi level. The hybridized Cr~$3d$ ($t_{2g}$) spin-down states
and the W~$5d$ ($t_{2g}$) spin-down states form a broad spin-down band with
the Fermi level lying in this band.  This is fully consistent with the
model presented in Fig.~\ref{Fig:models}. Furthermore, the band structure
calculation shows that La doping is expected to add electrons in the
spin-down band.

\subsection{Optical Measurements}

\begin{figure}[bt]
\centering{%
\includegraphics [width=0.9\columnwidth,clip=]{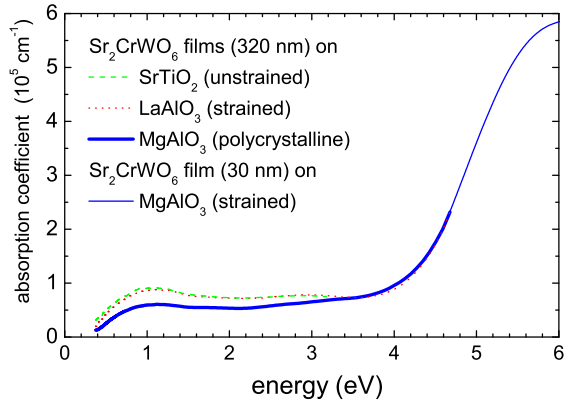}}
 \caption{Optical absorption coefficients of Sr$_2$CrWO$_6$ thin films calculated
from reflection and transmission measurements at room temperature.  The
films on SrTiO$_3$ (transparent in the range of 0.20-3.2\,eV) and LaAlO$_3$
(0.17-5.5\,eV) were epitaxial, and those on MgAl$_2$O$_3$ (0.22-6.5\,eV)
were polycrystalline. 30\,nm thin films were used for quantitative
measurements at high absorption coefficients above 4.6\,eV.}
 \label{Fig:Absorption}
\end{figure}

In order to verify the band structure calculations experimentally, we have
performed optical reflection and transmission measurements of
Sr$_2$CrWO$_6$ thin films with photon energies from 0.38\,eV to 7\,eV. The
SrTiO$_3$ substrates, on which the epitaxial quality of the films is high
(see Fig.~\ref{TEM}), are unfortunately transparent only in the range of
photon energies of 0.20-3.2\,eV. Therefore, we have also investigated
strained epitaxial Sr$_2$CrWO$_6$ films on LaAlO$_3$ substrates, which are
transparent from 0.17 to 5.5\,eV, and polycrystalline Sr$_2$CrWO$_6$ films
on MgAl$_2$O$_3$ substrates, which are transparent from to 0.22-6.5\,eV,
however have a larger lattice mismatch to Sr$_2$CrWO$_6$. On each
substrate, films with thicknesses of $d=30$, 80, and 320\,nm were
investigated. Almost identical optical absorption spectra were obtained for
the epitaxial films, suggesting that the observed optical features are
indeed related to the Sr$_2$CrWO$_6$ films, as summarized in
Fig.~\ref{Fig:Absorption}. The dominant optical features, which are an
absorption shoulder around 1\,eV, and a strong increase of the absorption
above 4\,eV, are also found for the polycrystalline samples grown on
MgAl$_2$O$_4$. The transmission $T$ at 4.6\,\,eV is around 0.1\% for the
films with $d=320$\,nm, and around 30\% for the films with $d=30$\,nm. The
error in the calculation of the absorption coefficient $\alpha =
\ln[(1-R)^2/T]/d $ with the reflection $R$ comes mostly from uncertainties
in $R/T$, which is sufficiently small for the 320\,nm films at lower
energies, and for the 30\,nm films at higher energies. For the films with
$d=320$\,nm, reflectivity oscillations with a period $p=0.77\pm 0.07$\,eV
were observed at photon energies of 1.5-3.5\,eV, indicating a refractory
index $ n = hc/(2\ d\ p)=2.5\pm 0.2$. This is consistent with the measured
reflection data and $R=(n-1)^2/(n+2)^2\approx 18\pm 4\%$ in this energy
range. As determined from Fourier transform infrared (FTIR) transmission
measurements, the films remain transparent down to 0.2\,eV, however, $R$ is
increasing significantly towards lower photon energies.

The optical measurements agree fairly well with the band structure
calculations. The increase of the absorption coefficient of Sr$_2$CrWO$_6$
above 4\,eV can most probably be attributed to a charge transfer transition
between the p-like spin-up and -down oxygen bands at $-3$\,eV below the
Fermi-level into the oxygen/metal bands at +1\,eV above the Fermi-level.
The absorption shoulder around 1\,eV coincides roughly with the energy gap
at the Fermi level in the spin-up band, and is therefore probably caused by
transitions from the Cr spin-up $t_{2g}$ states below the Fermi-level into
the oxygen/metal bands around +1\,eV above the Fermi-level. Unfortunately,
due to the substrate absorption below 0.2\,eV, no optical information is
available in the infrared region which would allow a more definitive
statement with respect to the density of states at the Fermi level and the
half-metallic character of Sr$_2$CrWO$_6$. However, from the refractory
index $n \approx 2.5$, one can estimate an optical conductivity of
$\sigma_{\text{opt}} = \alpha n c \epsilon_0 \approx \ =
630\,\Omega^{-1}$cm$^{-1}$ at 1\,eV. Comparing this result with recent
measurements of Sr$_2$CrReO$_6$ \cite{Kato:02}, one can classify
Sr$_2$CrWO$_6$ as a (very) bad half metal. This is also in agreement with
recent transport measurements \cite{Philipp:01}. We note, however, that the
transport data may be ambiguous, since the conductivity of the
Sr$_2$CrWO$_6$ thin film is comparable to that of the substrate. This is
for example the case for SrTiO$_3$ substrates, which obtain a conductive
surface layer at the reducing atmosphere of the thin film deposition
process \cite{Philipp:03}. That is, more conclusive transport data are
required to settle this issue.

\section*{Summary}

We have performed a detailed analysis of the structural, transport,
magnetic and optical properties of of the double perovskite $A_2$CrWO$_6$
with $A=\text{Sr, Ba, Ca}$. In agreement with band structure calculations
the double perovskite Sr$_2$CrWO$_6$ is a half-metal with a high
Curie-temperature above 450\,K. The measured saturation magnetization of
$1.11\,\mu_B/$f.u. is well below the optimum value of $2\,\mu_B/$f.u. due
to 23\% of Cr/W antisites. The large amount of antisites most likely is
caused by the similar ionic radii of the Cr$^{3+}$ and W$^{5+}$ ions
resulting in a low threshold for the formation of antisites.

The substitution of Sr by Ba and Ca in the double perovskite system
$A_2$CrWO$_6$ showed that the maximum Curie temperature is obtained for the
compound with a tolerance factor close to one. This is in agreement with a
large variety of data reported in literature on other double perovskite
systems $A_2BB^\prime$O$_6$. The fact that the maximum $T_C$ is obtained
for a tolerance factor close to one differs from the behavior of the doped
manganites where a maximum $T_C$ was found for a tolerance factor of about
0.93. The observation that a tolerance factor $f\simeq 1$ yields the
highest $T_C$ has been explained within a model, where ferromagnetism is
stabilized by the energy gain contributed by the negative spin polarization
on the nonmagnetic W-site due to hybridization of the Cr~$3d$ ($t_{2g}$)
and the W~$5d$ ($t_{2g}$) states. This hybridization is weakened by bond
angles deviating from $180^\circ$ or equivalently a tolerance factor
deviating from unity.

Electron doping of Sr$_2$CrWO$_6$ by partial substitution of Sr$^{2+}$ by
La$^{3+}$ was found to decrease both the Curie temperature and the
saturation magnetization. The decrease in the saturation magnetization was
found to be caused both by an increase in the amount of antisites and by
increasing band filling. Although the decrease of $T_C$ with increasing
doping can be explained qualitatively within the same model, it seems to be
in conflict with double exchange type models predicting an increase of
$T_C$ with increasing band filling. However, an unambiguous conclusion
cannot be drawn at present. The reason for that is the fact that on
increasing the doping level one also obtains an increasing amount of
disorder. In order to clarify this issue it is required to study systems
allowing for the variation of the doping level without changing disorder.

This work was supported by the Deutsche Forschungsgemeinschaft and the
Bundesministerium f\"{u}r Bildung und Forschung (project 13N8279). The authors
acknowledge fruitful discussions with M. S. Ramachandra Rao.

\end{document}